# Tunneling Plasmonics in Bilayer Graphene

Z. Fei[1], E. G. Iwinski[1], G. X. Ni[1], L. M. Zhang[1,3], W. Bao[4], A. S. Rodin[3], Y. Lee[4], M. Wagner[1], M. K. Liu[1,5], S. Dai[1], M. D. Goldflam[1], M. Thiemens[6], F. Keilmann[7], C. N. Lau[4], A. H. Castro-Neto[2,3], M. M. Fogler[1], D. N. Basov[1]

[1]Department of Physics, University of California, San Diego, La Jolla, California 92093, USA
[2]Graphene Research Centre, National University of Singapore, 117542, Singapore
[3]Department of Physics, Boston University, Boston, Massachusetts 02215, USA
[4]Department of Physics and Astronomy, University of California, Riverside, California 92521, USA
[5]Department of Physics, Stony Brook University, Stony Brook, New York, 11794, USA
[6]Department of Chemistry and Biochemistry, University of California, San Diego, La Jolla, California 92093, USA
[7]Ludwig-Maximilians-Universität and Center for Nanoscience, 80539 München, Germany

## Abstract

We report experimental signatures of plasmonic effects due to electron tunneling between adjacent graphene layers. At sub-nanometer separation, such layers can form either a strongly coupled bilayer graphene with a Bernal stacking or a weakly coupled double-layer graphene with a random stacking order. Effects due to interlayer tunneling dominate in the former case but are negligible in the latter. We found through infrared nano-imaging that bilayer graphene supports plasmons with a higher degree of confinement compared to single- and double-layer graphene, a direct consequence of interlayer tunneling. Moreover, we were able to shut off plasmons in bilayer graphene through gating within a wide voltage range. Theoretical modeling indicates that such a plasmon-off region is directly linked to a gapped insulating state of bilayer graphene: yet another implication of interlayer tunneling. Our work uncovers essential plasmonic properties in bilayer graphene and suggests a possibility to achieve novel plasmonic functionalities in graphene few-layers.

## Main Text

Quantum plasmonics is a rapidly growing field of research that involves the study of light-matter interactions in the quantum regime[1,2]. In particular, tunneling plasmonics explores the role of electron tunneling on the optical responses of coupled plasmonic nanostructures. Previous studies about tunneling plasmonics were focused on coupled metal nanoparticles within sub-nanometer separation[3-6]. The plasmonic resonance of the coupled system deviates from classical predictions where only Coulomb coupling is considered. In order to fully describe the plasmonic response, quantum tunneling of electrons between the nanoparticles has to be taken into account. Here we report experimental observation of novel plasmonic phenomena due to electron quantum tunneling between adjacent layers of graphene － a novel plasmonic material that carries infrared plasmons with high confinement, low loss and gate tunability[7-13]. Specifically, we observed a high plasmon confinement and effective plasmon-off state when two graphene layers are placed close to each other in a Bernal-stacking order. These effects are attributed to interlayer electron tunneling that plays a prominent role in graphene due to the reduced dimensionality and relatively low carrier density.

In order to study the plasmonic properties of coupled graphene layers, we performed infrared (IR) nano-imaging experiments using an antenna-based nanoscope (Supporting Information). As shown in Figure 1a, the metalized tip of the atomic force microscope (AFM) is illuminated by IR light thus generating strong near fields underneath the tip apex. These fields have a wide range of in-plane momenta $q$ therefore facilitating energy transfer and momentum bridging from photons to plasmons[7-13]. Our samples were fabricated by mechanical exfoliation of bulk graphite and then transferred to $SiO_2$/Si wafers. The thickness and stacking of the graphene layers were determined through a combination of optical microscopy, AFM, and Raman spectroscopy (Figure S1).

In Figure 1b,c,e, we show typical IR nano-imaging data taken at an excitation frequency of $\omega_{IR}$ = 883 cm$^{-1}$ (corresponding to a photon energy of 109 meV), where we plot the amplitude signal $s$ of the back-scattered radiation by the AFM tip (Supporting Information). We investigated samples where Bernal-type bilayer graphene (Figure 1d) that is adjacent to single- and double-layer graphene. The latter is in fact folded single-layer graphene with a random stacking order (Figure S1). We emphasize that it is critical to compare all three forms of graphene samples: single-layer, bilayer and double-layer graphene where they are adjacent to each other so that they share identical substrate and environmental conditions. This ensures a uniform carrier density due to the unintentional doping[7,12]: an assertion that we have confirmed by gating experiments (Figure 2).

Bright fringes in Figure 1 were observed close to the edges of all graphene layers. According to previous studies[8-10,12,13], such fringes are formed when surface plasmon waves launched by the tip interfere with those reflected by edges or defects. These alternating bright and dark fringes have a period of $\lambda_p/2$: one half of the plasmon wavelength. The fact that the periods of all the fringes shrinks with increasing excitation frequency verifies their plasmonic origin (Figure S2).

It is evident from Figure 1b,c,e that the plasmon fringes of the bilayer graphene are slightly narrower than those of single-layer graphene[13], while plasmon fringes of double-layer graphene appeared to be much wider [12]. For the purpose of quantitative comparison, we plot in Figure 1f the line profiles taken perpendicular to edges of all graphene layers along the dashed lines shown in Figure 1e. The plasmon wavelengths, which are read off directly from the profiles by doubling the fringe period, are 150, 188, and 243 nm for bilayer, single-layer and double-layer graphene, respectively. We wish to stress that the $\lambda_p^{BLG} < \lambda_p^{SLG} < \lambda_p^{DLG}$ inequality is a common observation for all our graphene samples so long as these graphene layers are adjacent to each other. Therefore, the plasmon confinement factor $C = \lambda_{IR}/\lambda_p$ for surface plasmons in bilayer graphene surpasses those of adjacent single- and double-layer graphene. Here $\lambda_{IR} = 1/\omega_{IR}$ is the wavelength of excitation IR light.

We posit that the dramatic difference between $\lambda_p$ in bilayer and double-layer graphene stems from their distinct mechanisms of interlayer coupling. Unlike Bernal-type bilayer graphene, the top and bottom layers of double-layer graphene are stacked randomly and may also be separated by occasional surface deposits (Figure 1d). As a result, interlayer tunneling in double-layer graphene (green dashed arrow in Figure 1d) is strongly suppressed compared to that of bilayer graphene (green solid arrow in Figure 1d). Therefore double-layer graphene can be treated as two independent graphene planes coupled only by the Coulomb interaction[12,14,15]. Within this scenario, we can estimate the ratio between $\lambda_p$ of double-layer graphene with sub-nanometer interlayer separation and single-layer graphene to fall in the range $1 < \lambda_p^{DLG}/\lambda_p^{SLG} < \sqrt{2}$ (Supporting Information)[12].

Indeed, the experimental values of $\lambda_p^{DLG}/\lambda_p^{SLG}$ obtained from the two double-layer graphene samples shown in Figure 1c,e are around 1.4 and 1.3, respectively, consistent with the above inequality. The analysis of bilayer graphene is more complicated since it necessitates a proper account of the interlayer tunneling: an effect responsible for dramatic changes in the electronic structure[15-17]. The direct outcome of the tunneling is that the plasmon wavelength of bilayer graphene is always smaller than that of equally doped single- and double-layer graphene as will be discussed in detail below (Figure 3a).

So far, we discussed the case of unintentional doping. However, electrostatic tuning of graphene is readily attainable in gated structures. Through back-gating, we were able to explore the key aspects of tunneling plasmonics on bilayer graphene plasmons by tuning both the Fermi energy as well as interlayer doping asymmetry[16,18,19]. In Figure 2, we show a selected dataset of gate-dependent near-field images of a graphene sample containing both single-layer and bilayer graphene. Unless otherwise specified, we discuss mainly the voltage difference $V_g - V_{CN}$, where $V_g$ is the back gate voltage and $V_{CN}$ is voltage for the charge neutrality point (CNP). When $V_g = V_{CN}$, the entire sample becomes charge neutral thus no plasmon fringe is observed (Figure 2c). At high doping regime (Figure 2a,f), we observed plasmons in both single-layer and bilayer graphene. However, when the sample is weakly doped (Figure 2b,d), dramatic differences appear: the plasmon fringes in single-layer graphene are clearly visible, but those in bilayer graphene are not observed.

In Figure 3a we plot the complete voltage dependence of $\lambda_p$ for bilayer graphene (blue dots, labeled as BLG-1) and single-layer graphene (black dots) extracted from gate-dependent near-field images including those in Figure 2. In addition, we also plot data points from another bilayer graphene sample (red dots, labeled as BLG-2, see Figure S4). From Figure 3a, one can see that $\lambda_p$ for all the three samples show obvious ambipolar gate voltage dependence: $\lambda_p$ increases with either higher electron or hole density[9]. However, as graphene approaches the CNP, rather striking differences between single-layer and bilayer graphene emerge. For single-layer graphene, $\lambda_p$ drops to zero only in the immediate proximity to the CNP, as expected. In contrast, for bilayer graphene, we observed an extended voltage range where there are no detectable mid-IR plasmons. Such a plasmon-off region can also be seen in Figure S3, where we plot the color fringe profiles taken perpendicular to the edges of single-layer and bilayer graphene (along the green and blue dashed lines in Figure 2a). These profiles are extracted directly from the near-field images acquired at various gate voltages (including those in Figure 2) and together they show visually the evolution of plasmons with gate voltages as well as the plasmon-off region. The width of the plasmon-off region $W$ determined from our data is 45 ±5 V (blue double arrow in Figure 3a) and 60 ±5 V (red double arrow in Figure 3a) for BLG-1 and BLG-2, respectively. The uncertainty is primarily due to the spatial resolution (~20 nm) of our technique.

In order to account for the gate-dependence of $\lambda_p$, we performed tight-binding calculations of bilayer graphene taking into consideration the effect of interlayer tunneling $\gamma \approx 0.4$ eV on the electronic structure[16]. The modeling has one adjustable parameter $\delta n_0$ that defines the impurity-induced interlayer asymmetric doping, which corresponds to the carrier density of the top ($\delta n_0$) and bottom ($-\delta n_0$) layers at the CNP. When applying back-gate voltages, more carriers will be injected into the bottom layer compared to the top layer due to the screening of the bottom layer. Therefore, the bandgap $\Delta$ of bilayer graphene (inset of Figure 3b) that is roughly proportional to $|n_{top} - n_{bot}|$ will systematically evolve with gate voltages (Figure 3b)[23]. Given a trial $\delta n_0$, we determined the band gap $\Delta(V_g)$ and Fermi energy $E_F(V_g)$ of bilayer graphene in the experimental range of gate voltages and then computed the gate-dependent optical conductivity using the Kubo formula[16]. This latter computation is sufficient to evaluate $\lambda_p$ in bilayer graphene (Supporting Information) at any excitation frequency. The best agreement with our data for BLG-1 and BLG-2 is obtained assuming $\delta n_0 = 2.5 \times 10^{11}$ cm$^{-2}$ and $\delta n_0 = 9.1 \times 10^{11}$ cm$^{-2}$. The corresponding bandgaps of BLG-1 and BLG-2 at the CNP ($\Delta_{CN}$) are about 24 meV and 100 meV, respectively. The finite $\Delta_{CN}$ is enabled by initial high impurity-doping of our samples exposed to ambient environment.

For comparison, we also show in Fig. 3a the calculation results for double-layer graphene (red and blue dashed curves) and bilayer graphene with parabolic bands (green curve). They all deviate largely from the experimental data points. For double-layer graphene, the calculated $\lambda_p$ is much bigger than that of BLG-1 and BLG-2 and there appears to be more than one minimum in the gate-dependent $\lambda_p$ curves of double-layer graphene due to the different Dirac-point energies of the top and bottom layers. In the case of the parabolic-band model, it is known to be a good approximation of gapless bilayer graphene at low doping regime (the bands become linearized at high doping regime, see Section 7 of the Supporting Information for more discussions). Here close to the CNP, we also see a plasmon-off region in the modeled $\lambda_p$ curve assuming parabolic bands (green curve), but the size of the plasmon-off region is smaller than that of BLG-1 and BLG-2. The difference in size is attributed to interlayer asymmetric doping or bandgap opening (discussed in detail below) that is not considered in the parabolic-band model.

With the help of the tight-binding model (Figure 3a), we were able to determine the width of the plasmon-off region $W$ at various $\delta n_0$. In Figure 4a, we plot the $W(\delta n_0)$ dependence at $\omega_{IR} = 883$ cm$^{-1}$ (black curve) and $\omega_{IR} = 100$ cm$^{-1}$ (green curve) that is in the terahertz (THz) range – another important regime for graphene plasmonics[24,25]. We found that $W$ scales monotonically with $\delta n_0$. Therefore, $\delta n_0$ can be readily estimated by comparing the measured $W$ to the theoretical curve. Note that $W(\omega_{IR} = 883$ cm$^{-1})$ starts with a finite value and is clearly larger than $W(\omega_{IR} = 100$ cm$^{-1})$. To understand this frequency dependence of $W$, we show in Figure 4b-f the voltage- and frequency-dependent maps of the imaginary part of the calculated optical conductivity $\sigma_2(V_g-V_{CN}, \omega_{IR})$ for BLG-1 and BLG-2 as well as their single-layer (labeled as SLG) and double-layer graphene (labeled as DLG-1 and DLG-2) counterparts. The asymmetric doping parameter $\delta n_0$ is assumed to be the same for BLG-1 and DLG-1, and for BLG-2 and DLG-2. As explained in the Supporting Information and Ref. 8, well-defined surface plasmons may exist only when the complex conductivity of graphene $\sigma = \sigma_1 + i\sigma_2$ is predominantly imaginary: $\sigma_2 \gg \sigma_1$. In this regime, the plasmon wavelength $\lambda_p$ is approximately proportional to $\sigma_2$. Therefore, the red regions of the color maps in Figure 4b-f where $\sigma_2 > 0$ and also $\sigma_2 \gg \sigma_1$ (Figure S5) display the plasmon-on state. The white and blue regions where $\sigma_2 \leq 0$ are at the plasmon-off state. As a result, the width of the plasmon-off territory $W$ at every given frequency can be approximately determined from Figure 4b-f. The magnitude of $W$ for BLG-1 and BLG-2 at $\omega_{IR} = 883$ cm$^{-1}$ are indicated by the blue and red arrows, respectively.

As can be seen in Figure 4b,c, the width of the plasmon-off region $W$ increases with frequency for both BLG-1 and BLG-2. We first focus on the low frequency regime (e.g. THz region) where the Drude response dominates. Here $W$ of both single-layer and double-layer graphene vanishes, making it impossible to turn off plasmons effectively – a dilemma similar to that faced by graphene field-effect transistors. In contrast, for bilayer graphene with finite $\delta n_0$, the bandgap opens up close to the CNP and the Fermi level falls inside the gap (Figure 3b). As a result, an insulating region occurs between the two vertical dashed lines in Figure 4b,c[26,27]. Within such a region, plasmons are turned off completely over the entire frequency range as clearly shown in the dispersion diagrams (Figure S6). Therefore such a gapped insulating region forms the central part of the plasmon-off region. The width of this insulating region $W_i$ is roughly equal to $W$ at $\omega = 100$ cm$^{-1}$. Therefore plasmon measurements at the THz frequency regime serve a similar role as transport measurements in determining the gap size of bilayer graphene. According to our analysis, the width of this insulating region $W_i$ scales with $|\delta n_0|$ and BLG-2 has a larger $W_i$ than BLG-1 (Figure 4a).

Now let's look at the mid-IR regime close to our excitation frequencies. Here the plasmon-off region of bilayer graphene becomes wider than the insulating region (Figure 4b,c). This is because more carriers have to be injected into bilayer graphene to elevate plasmons to the probing IR frequency and to lift plasmons off the Landau damping regime by interband transitions (green arrows in the inset of Fig. 3b). These interband transitions reduce $\sigma_2$ to zero and even negative values[20] (blue and white regions in Figure 4b-f) and thus extends the plasmon-off region ($\sigma_2 \leq 0$) further away from the CNP. The width of the extended plasmon-off region due to interband transitions exceeds 30 V for both BLG-1 and BLG-2 at $\omega_{IR} = 883$ cm$^{-1}$. Within the extended plasmon-off region, plasmons are turned off at our excitation frequency but still exist at THz frequencies as clearly shown in the dispersion diagrams in Figure S6. Note that the interband transitions are also responsible for the plasmon-off regions in gapless bilayer graphene with parabolic bands and even single-layer graphene at $\omega_{IR} = 883$ cm$^{-1}$ (Figure 3a). As a result, they share the same size in energy units as shown in Figure S7b in the Supporting Information where we plot $\lambda_p$ versus $E_F$.

Our work explores tunneling plasmonics of coupled graphene layers by comparing Bernal-type bilayer graphene with single-layer and randomly-stacked double-layer graphene. The strong interlayer tunneling endows bilayer graphene with novel attributes of plasmonic behaviors including the ability to efficiently turn the surface plasmons on and off by gating and an enhanced confinement compared to its single- and double-layer graphene counterparts. These properties augmented with inherent tunability, make bilayer graphene a promising material for practical implementation of plasmonic transistors and switches in future plasmonic circuitry[28]. Future studies may explore tunneling plasmonics in other forms of graphene bilayers, such as AA-stacked and twisted bilayer graphene. In the latter material, plasmons could be modified by periodic Moiré potential originated from lattice misorientation[29]. Our work opens up a new frontier in the study of the plasmonics in graphene where quantum physics plays a crucial role[30].

## Associated Content

Supporting Information. Supporting experimental data and theory details. This material is available free of charge via the Internet at http://pubs.acs.org.

## Author Information

Corresponding Author
*Email: (Z.F) zfeiphy@gmail.com

## Financial Interests

The authors declare the following competing financial interest(s): F. K. is cofounder of Neaspec, producer of the s-SNOM used in this study. All other authors declare no competing financial interests.

## Acknowledgment

Authors acknowledge support from ONR Grant N00014-15-1-2671 and UC Lab Fees Research Program Award #237789. The development of scanning plasmon interferometry is supported by DOE-BES and ARO. D.N.B. is the Moore Investigator in Quantum Materials Grant GBMF4533. A.H.C.N. acknowledges NRF-CRP Grant R-144-000-295-281.

# Figures

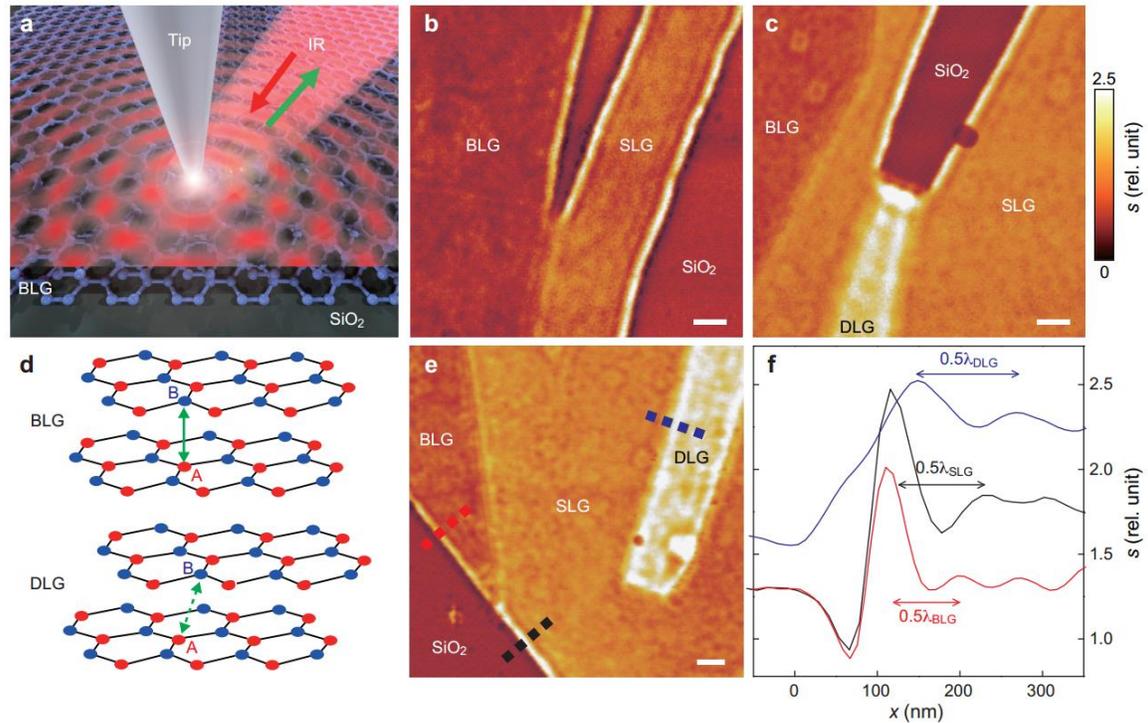

**Figure 1.** Infrared nano-imaging revealing plasmons on different graphene layers. **a**, Schematics of our infrared nano-imaging experiment. **b,c,e**, Infrared near-field imaging data of graphene samples containing single-layer graphene (SLG), bilayer graphene (BLG) and double-layer graphene (DLG) taken at $\omega_{IR} = 883$ cm$^{-1}$ (corresponding to a photon energy of 109 meV). Scale bars, 200 nm. **d**, Illustrations of lattice structures of Bernel-type bilayer graphene and randomly stacked double-layer graphene. The green solid arrow illustrates the interlayer electron tunneling between sublattices A (red atoms) and B (blue atoms) in bilayer graphene. The green dashed arrow illustrates the strongly suppressed interlayer tunneling in double-layer graphene. **f**, Line profiles taken perpendicular to the edges of single-layer (black), bilayer (red) and double-layer (blue) graphene of the sample shown in **e**. These profiles are averaged over a distance of 100 nm along the edges to reduce noise. The double-sided arrows mark the width of half the plasmon wavelength. The plotted quantity $s$ in **b,c,e,f** is the near-field amplitude normalized to that of SiO$_2$.

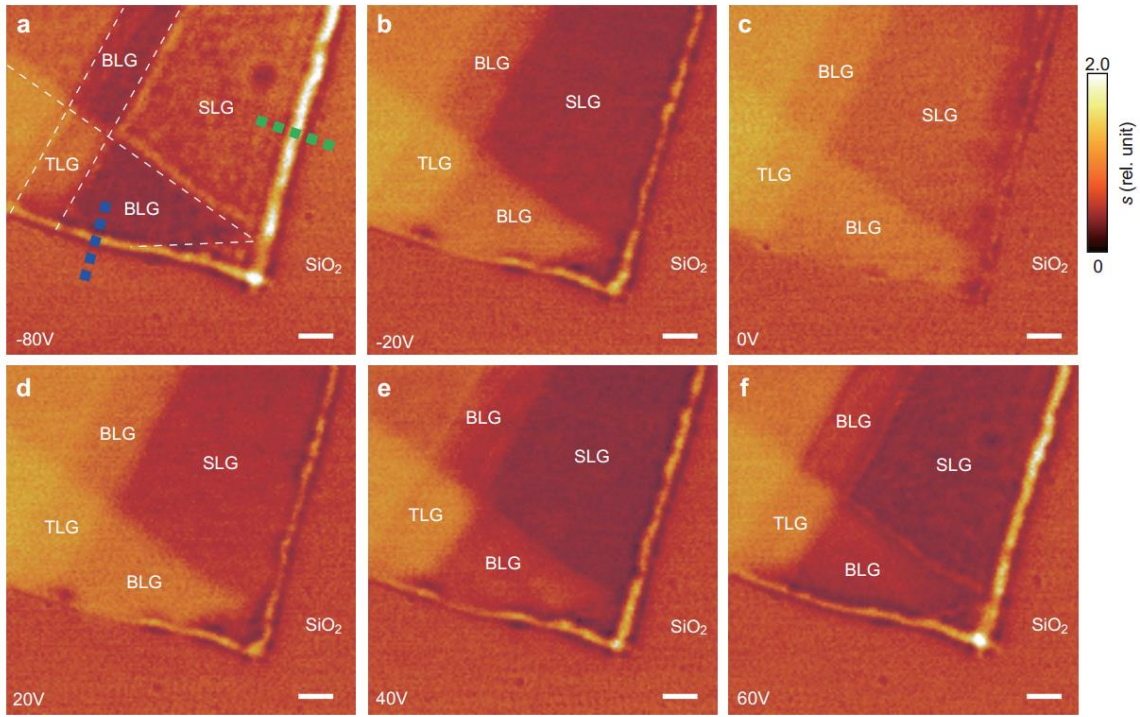

**Figure 2.** Infrared nano-imaging of single-layer and bilayer graphene under back gating. **a-f**, Infrared near-field images of a graphene sample containing bilayer graphene (BLG), single-layer graphene (SLG) and trilayer graphene (TLG) taken at an excitation wavelength $\omega_{IR} = 883$ cm$^{-1}$ under various gate voltages $V_g - V_{CN}$. The plotted quantity $s$ is the near-field amplitude normalized to that of SiO$_2$. Here gate voltages were applied on the silicon side, so positive $V_g - V_{CN}$ will induce electron doping to graphene samples. Green and red dashed lines in **a** marks the edges where we measure the plasmon wavelength of single-layer and bilayer graphene, respectively. The white dashed lines mark the boundary between different graphene layers. Scale bars, 200 nm.

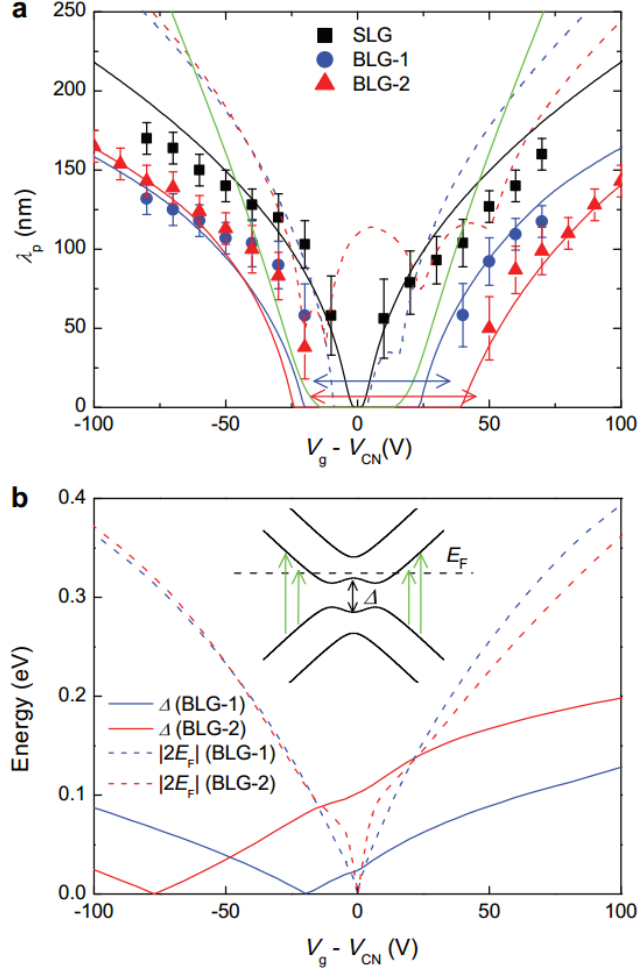

**Figure 3.** The plasmon-off region of bilayer graphene. **a**, Plasmon wavelength taken from the nano-imaging data of single-layer graphene (SLG) and two bilayer graphene samples (BLG-1 and BLG-2). The SLG (black dots) and BLG-1 (blue dots) data points were extracted from the profiles taken perpendicular to the edges of single-layer and bilayer graphene in Figure 2 (along dashed color lines as illustrated in Figure 2a). The BLG-2 data points (red dots) were obtained from near-field images of a different sample (Figure S4). The black, blue and red solid curves are theoretical calculations about SLG, BLG-1 ($\delta n_0 = 2.5 \times 10^{11}$ cm$^{-2}$) and BLG-2 ($\delta n_0 = 9.1 \times 10^{11}$ cm$^{-2}$), respectively. The green solid curve is the calculated plasmon wavelength of bilayer graphene in the parabolic-band approximation. The blue and red dashed curves are theoretical calculations of double-layer graphene assuming $\delta n_0 = 2.5 \times 10^{11}$ cm$^{-2}$ (DLG-1) and $\delta n_0 = 9.1 \times 10^{11}$ cm$^{-2}$ (DLG-2), respectively. **b**, Calculation of the bandgap $\Delta$ (solid curves) and Fermi energy ($\times 2$, dashed curves) at various gate voltages for BLG-1 (blue) and BLG-2 (red), respectively. Inset plots a typical example of the band structure of gapped bilayer graphene. For the purpose of illustration, a very large bandgap ($\Delta = \gamma$) is used here. The black dashed line here denotes the Fermi level and the green arrows illustrate the interband transitions above $2E_F$.

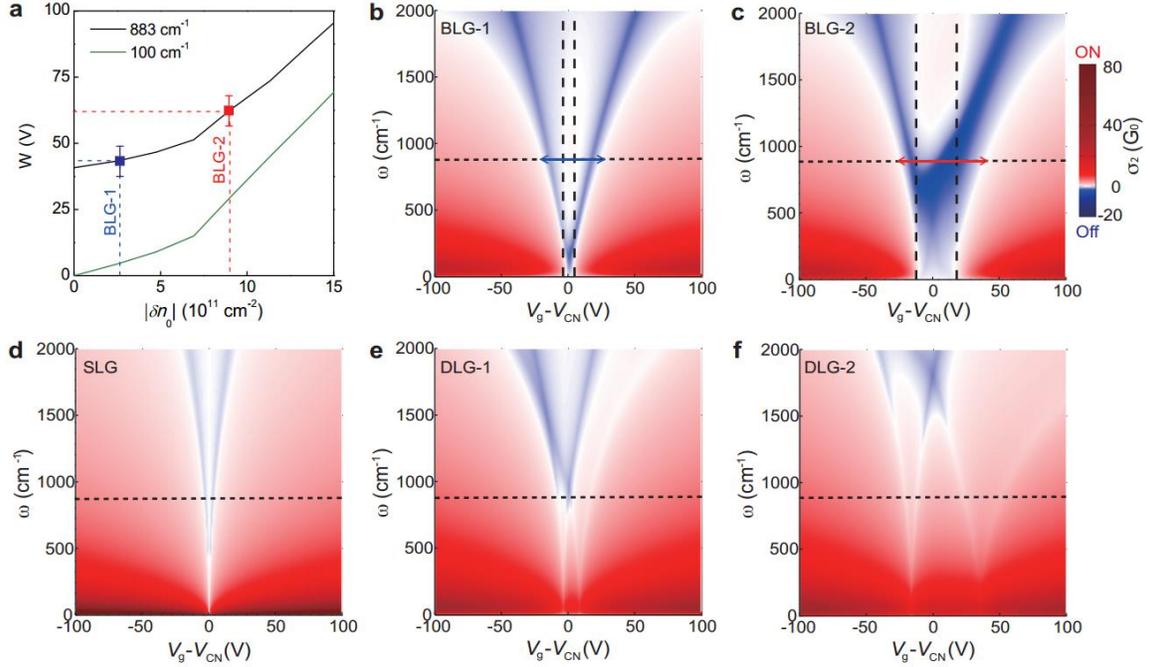

**Figure 4.** Theoretical description of the plasmon-off region of bilayer graphene. **a,** Calculated $W(\delta n_0)$ dependence curves at two excitation frequencies. The blue and red dots mark the positions of the two bilayer graphene samples investigated (BLG-1 and BLG-2). **b-f,** Calculated Voltage- and frequency- dependent maps of the imaginary part of the optical conductivity $\sigma_2(V_g\text{-}V_{CN}, \omega_{IR})$ of BLG-1, BLG-2, SLG, DLG-1 and DLG-2 as detailed in the main text. Here, the parameter $\delta n_0$ is set to be $2.5 \times 10^{11}$ cm$^{-2}$ and $9.1 \times 10^{11}$ cm$^{-2}$ for BLG-1 (DLG-1) and BLG-2 (DLG-2), respectively. The red regions of these color maps correspond to plasmon-on state while the blue and white regions correspond to the plasmon-off state. The horizontal dashed line denotes our excitation frequency $\omega_{IR} = 883$ cm$^{-1}$ (corresponding to a photon energy of 109 meV). The two vertical dashed lines in **b** and **c** define the region of the gap-induced insulating state in bilayer graphene. The blue and red arrows in **b** and **c** mark the width of the plasmon-off region at our excitation frequency for BLG-1 and BLG-2, respectively. The unit $G_0$ for the conductivity map is $\pi e^2/2h$.

# Supporting Information for
# "Tunneling Plasmonics in Bilayer Graphene"

Z. Fei[1], E. G. Iwinski[1], G.-X. Ni[1,2], L. M. Zhang[1,3], W. Bao[4], A. S. Rodin[3], Y. Lee[4], M. Wagner[1], M. K. Liu[1,5], S. Dai[1], M. D. Goldflam[1], M. Thiemens[6], F. Keilmann[7], C. N. Lau[4], A. H. Castro-Neto[2,3], M. M. Fogler[1], D. N. Basov[1]

[1]Department of Physics, University of California, San Diego, La Jolla, California 92093, USA
[2]Graphene Research Centre, National University of Singapore, 117542, Singapore
[3]Department of Physics, Boston University, Boston, Massachusetts 02215, USA
[4]Department of Physics and Astronomy, University of California, Riverside, California 92521, USA
[5]Department of Physics, Stony Brook University, Stony Brook, New York, 11790, USA
[6]Department of Chemistry and Biochemistry, University of California, San Diego, La Jolla, California 92093, USA
[7]Ludwig-Maximilians-Universität and Center for Nanoscience, 80539 München, Germany

**List of contents**

**Figure S7**. Experimental and modeling plasmon wavelength ($\lambda_p$) versus Fermi energy ($E_F$)

1. **Sample fabrication and characterization**

All our single-layer, bilayer and double-layer graphene samples were obtained by mechanical exfoliation of bulk graphite crystals and then transferred to silicon wafers with 285 nm thermal $SiO_2$ on the top. In order to determine the thickness and stacking of the graphene samples, we employed optical microscopy, Raman spectroscopy, and atomic force microscopy (AFM) to characterize all our graphene samples. In Figure S1, we present reprehensive characterization results of the samples shown in Figure 1b and 1c in the main text. The optical microscope images in Figures S1a and S1d clearly show thickness contrast of the graphene layers including single-layer graphene (labeled as SLG), Bernal-type bilayer graphene (labeled as BLG) and randomly stacked double-layer graphene (labeled as DLG) in the two samples. The double-layer graphene sample is in fact folded single-layer graphene as clearly shown in Figure S1d where a triangle shape single-layer graphene (originally sitting in the area marked with yellow dashed triangle) was folded to the other side of the sample during exfoliation. Raman spectroscopy data taken on different parts of the samples are shown in Figure S1b,e. The 2D peaks at around 2700 cm$^{-1}$ show clear signatures about both the thickness and stacking of graphene layers. For example, both SLG and DLG have symmetric single-Lorentzian-component 2D peaks while the Bernal-stacking bilayer graphene has a wider 2D peak with a featured four-Lorentzian-component shape. White dashed squares in Figures S1a and S1d are the areas we chosen for infrared (IR) nano-imaging (Figure 1b,c in the main text). Simultaneously collected AFM images in these areas are given in Figure S1c,f, where white dashed lines mark the boundaries between different graphene layers.

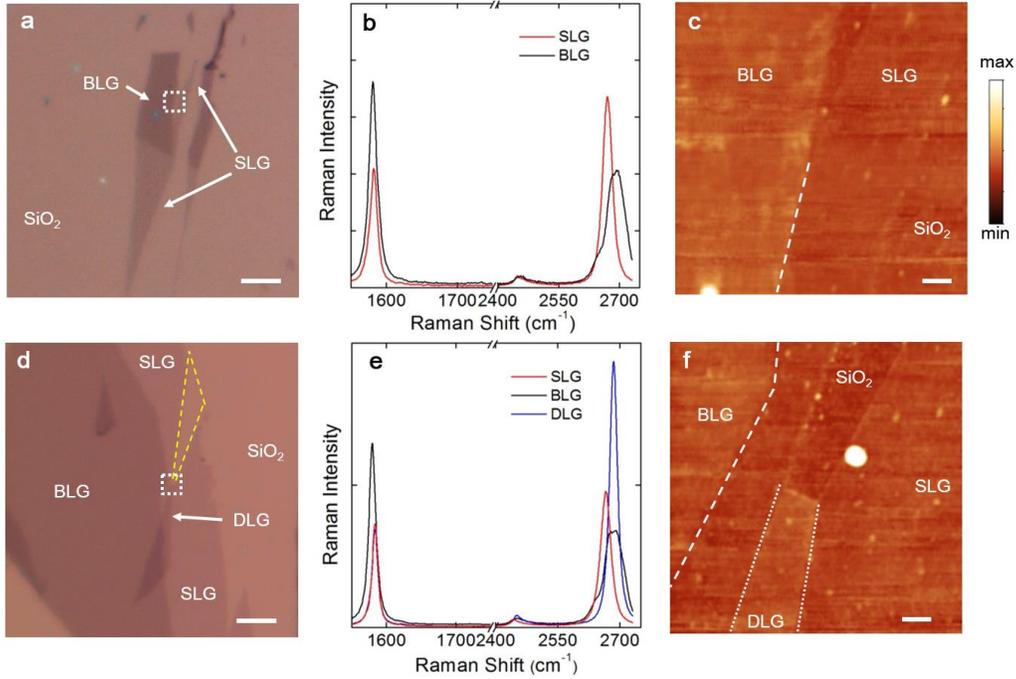

**Figure S1.** Thickness and stacking determination of graphene layers. **a-c**, Optical microscopy, Raman spectroscopy and AFM characterization of graphene sample shown in Figure 1b of the main text. **d-f,** Optical microscopy, Raman spectroscopy and AFM characterization of graphene sample shown in Figure 1c of the main text. Dashed squares in **a** and **d** mark the areas we chosen for infrared nano-imaging. Dashed lines in **c** and **f** mark the boundary between single-layer, bilayer and double-layer graphene.

2. **Infrared nano-imaging experiments**

The infrared (IR) nano-imaging experiments introduced in the main text were performed using a scattering-type scanning near-field optical microscope (s-SNOM). Our s-SNOM is a commercial system (neaspec.com) equipped with continuous wave mid-IR quantum cascade lasers (daylightsolutions.com) and $CO_2$ lasers (accesslaser.com). The unit for the IR frequency used in the work is wavenumber (cm$^{-1}$). The typical laser frequency used in the work is $\omega = 883$ cm$^{-1}$, corresponding to a photon energy of 109 meV. The s-SNOM is based on an atomic force microscope (AFM) operating in the tapping mode with a tapping frequency of ~270 kHz and tapping amplitude of ~50 nm. A pseudo-heterodyne interferometric detection module is implemented in our s-SNOM to extract both scattering amplitude $s$ and phase $\psi$ of the near-field signal. In the current work, we discuss mainly the amplitude part of the signal that is sufficient to determine the plasmon wavelength. In order to subtract the background signal, we demodulated the near-field signal at the $n^{th}$ harmonics of the tapping frequency ($n = 3$ in the current work). In all the displayed near-field images, we plotted the near-field amplitude normalized to that of the $SiO_2$ substrate. Our IR nano-imaging experiments were performed under ambient conditions and in an atmospheric environment.

## 3. Supporting infrared nano-imaging data

Supporting IR nano-imaging data are given in Figures S2, S3 and S4.

In Figure S2, we show the AFM topography image (Figure S2a) and frequency-dependent IR nano-imaging data (Figure S2b-d) of a graphene sample containing single-layer, bilayer and double-layer graphene. One can see from Figure S2a that the thicknesses of bilayer graphene and double-layer graphene are roughly the same, both of which have a step of less than 0.5 nm above single-layer graphene. The near-field nano-imaging data shown in Figure S2b-d are taken with excitation frequencies of $\omega_{IR}$ = 883 $cm^{-1}$, 903 $cm^{-1}$ and 923 $cm^{-1}$, respectively. Note that Figure S2b is a replot of Figure 1e in the main text. Clearly from Figure S2b-d, the bright fringes close to the edges of single-layer, bilayer and double-layer graphene become narrower and weaker with increasing excitation frequency. The narrowing of the fringe width is due to the decrease of the plasmon wavelength with increasing frequency, which is consistent with the plasmonic origin of these fringes[1,2]. The weakening of the intensity of these fringes indicates that our near-field probe is more sensitive to plasmons with wavelengths above 200 nm. When the plasmon wavelength drops towards 100 nm at higher excitation frequencies, the fringe intensity also decreases.

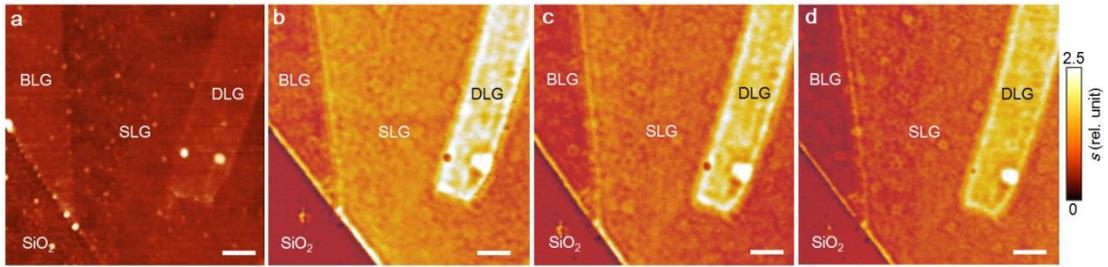

**Figure S2.** Infrared nano-imaging of graphene layers at various frequencies. **a**, AFM characterization of a graphene sample (also shown in Figure 1e of the main text) containing single-layer graphene (SLG), bilayer graphene (BLG) and double-layer graphene (DLG). b-d, Infrared nano-imaging data of the sample in **a** at 883 $cm^{-1}$, 903 $cm^{-1}$ and 923 $cm^{-1}$, respectively. Scale bars, 400 nm.

In Figure S3, we plot the line cuts across the plasmon fringes of single-layer and bilayer graphene in Figure 2 of the main text (along the green and blue dashed lines in Figure 2a). Each line cut, which appears as a horizontal stripe in Figure S3, is taken from a near-field image at a particular gate voltage. Together, all the line cuts form a position- and voltage-dependent map of plasmon fringes. From Figure S3, we can compared visually single-layer (SLG, Fig. S3a) and bilayer graphene (BLG-1, Fig. S3b). A number of observations can be seen directly from Figure S3. First, we notice that both SLG and BLG-1 show ambipolar gating dependence. Second, the plasmon fringes of BLG-1 is narrower compared to SLG indicating a smaller plasmon wavelength of BLG-1. Third,

BLG-2 has a much wider plasmon-off region compared to SLG as indicated by the double-sided arrows in Figure S3.

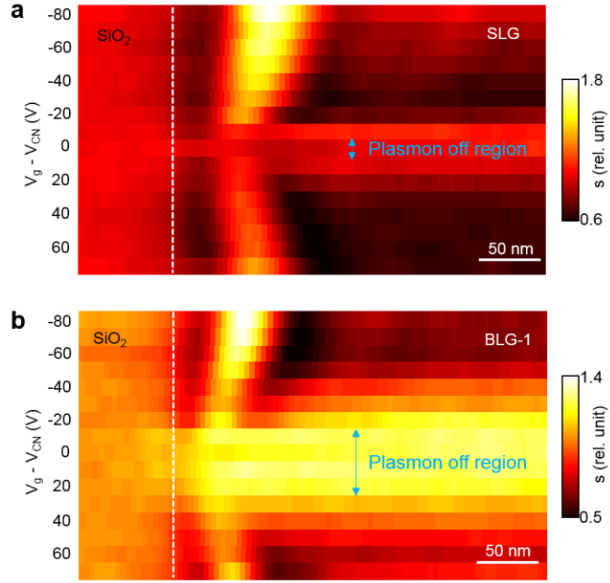

**Figure S3. a,b,** Gate-dependent plasmon fringes of single-layer (SLG) and bilayer graphene (BLG-1). There fringes are taken perpendicular to the edges of single-layer and bilayer graphene (along green and blue dashed lines in Figure 2a) from gate-dependent near-field images as exemplified in Figure 2.

In Figure S4, we show the gate voltage dependent IR nano-imaging data of the sample BLG-2, from which we extracted the gate-dependent wavelength data points (red triangles) in Figure 3a of the main text. From Figure S4, one can see strong plasmon fringes close to the edge when the bilayer graphene sample is highly electron or hole doped (Figure S4a,b). The plasmon fringes become narrower and weaker when the doping of the bilayer graphene sample decreases (Figure S4c,f) and are eventually barely seen when the sample is closer to the charge neutrality point (Figure S4d,e). In addition, we also found that the plasmon fringe intensity at the electron doping side ($V_g - V_{CN} > 0$) is clearly weaker than the hole doping side. This is probably because the bandgap of BLG-2 at the electron doping side is much bigger than the hole side (Figure 4b in the main text), which leads to a higher plasmon damping.

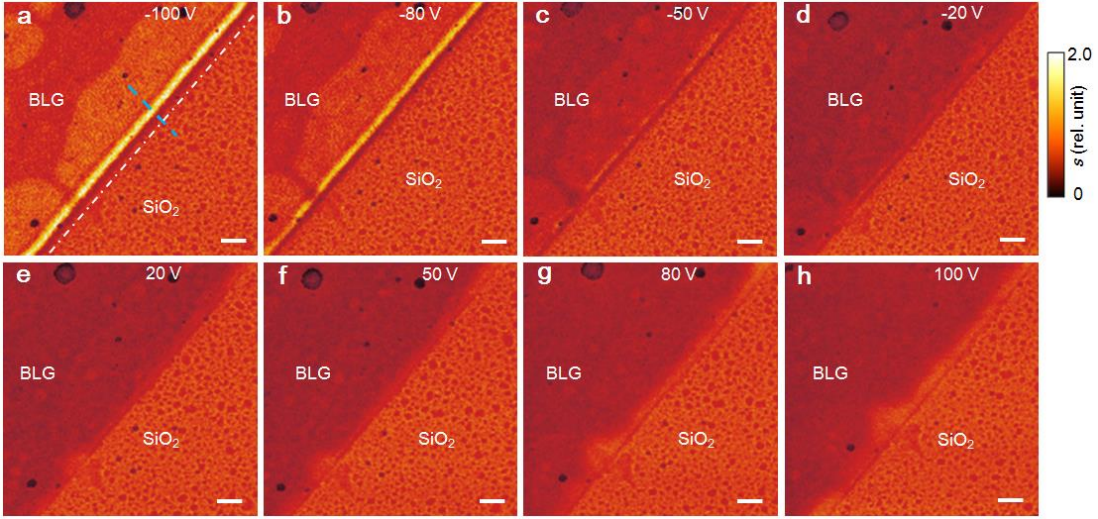

**Figure S4.** Gate voltage dependence of surface plasmons in the sample BLG-2. **a-f**, Infrared nano-imaging data of the sample BLG-2 taken at an excitation wavelength $\omega_{IR}$ = 883 cm$^{-1}$ under various voltages $V_g - V_{CN}$ from -100 to 100 V. Here gate voltages were applied on the silicon side, so positive $V_g - V_{CN}$ will induce electron doping to graphene samples. The white dashed line in **a** marks the edge of BLG. Scale bars, 200 nm.

## 4. Calculation of the optical conductivity

The optical conductivity of single-layer graphene $\sigma_{SLG}(\omega)$ used in our calculations is obtained with the Random Phase Approximation method[2,3]. The optical conductivity of parabolic-band approximated bilayer graphene $\sigma_{parabol}(\omega)$ can be written analytically as a sum of intra-band and inter-band terms:

$$\sigma_{parabol}(\omega) = \sigma_{parabol}^{Intra}(\omega) + \sigma_{parabol}^{Inter}(\omega) = \frac{2e^2 E_F}{\pi \hbar} \frac{i}{\omega + i\Gamma} + \frac{e^2}{2\hbar}\Theta(\omega - 2E_F).$$

Here $\omega$ has an energy unit and the step function $\Theta(\omega - 2E_F)$ can be replaced with a realistic form considering thermal broadening:

$$\Theta(\omega - 2E_F) \Rightarrow [\tanh(\frac{\omega + 2E_F}{4k_B T}) + \tanh(\frac{\omega - 2E_F}{4k_B T})] - \frac{i}{2\pi}\ln\left[\frac{(\omega + 2E_F)^2}{(\omega - 2E_F)^2 + (2k_B T)^2}\right].$$

The phenomenological scattering energy $\Gamma$ is set to be 100 cm$^{-1}$ and the thermal energy $k_B T$ is about 206 cm$^{-1}$ at ambient conditions. The Fermi energy has a form of $E_F = \hbar^2 \pi |n|/(2m_{eff})$ for bilayer graphene with simplified parabolic bands (effective mass $m_{eff} \approx 0.033 m_e$).

Randomly-stacked double-layer graphene is treated approximately as two independent single-layer graphene with a finite spacing. In our particular case, the separation between two graphene layers is within 1 nm far smaller than the plasmon wavelength, so we can treat the double-layer graphene as a two-dimensional film. Therefore the optical conductivity of double-layer graphene can be written as $\sigma_{DLG}(\omega) = \sigma_{top}(\omega) + \sigma_{bot}(\omega)$, where $\sigma_{top}(\omega)$ and $\sigma_{bot}(\omega)$ are the optical conductivities of the top and bottom graphene layers.

As introduced in Refs. 1 and 2, the plasmon dispersion of a two-dimensional film can be written as $q_p = i\omega\kappa(\omega)/2\pi\sigma(\omega)$ under the long-wavelength approximation. Here $\kappa(\omega) = [1+\varepsilon_{sub}(\omega)]/2$ is the effective dielectric function for the environment of graphene at the interface between air and substrate, $\sigma(\omega) = \sigma_1(\omega) + i\sigma_2(\omega)$ is the optical conductivity of the film. According to the dispersion equation, in order to observe well-defined plasmons, the optical conductivity should be predominantly imaginary ($\sigma_2 \gg \sigma_1$). In this case, the plasmon wavelength $\lambda_p$ is roughly proportional to $\sigma_2$. Within the Drude regime, we have $\lambda_p^{SLG} \propto \sigma_2 \propto \sqrt{n_{SLG}}$, where $n_{SLG}$ is the carrier density of single-layer graphene. While for double-layer graphene, the plasmon wavelength is $\lambda_p^{DLG} \propto \sqrt{n_{top}} + \sqrt{n_{bot}}$. Here, $n_{top}$ and $n_{bot}$ is the carrier density of the top and bottom layer of double-layer graphene. Therefore, the ratio between $\lambda_p$ of double- and single-layer graphene scales as $\lambda_p^{DLG}/\lambda_p^{SLG} \approx (\sqrt{n_{top}} + \sqrt{n_{bot}})/\sqrt{n_{SLG}}$. As discussed above, when single-layer graphene is adjacent to double-layer graphene, $n_{SLG} \approx n_{top} + n_{bot}$ due to their identical environmental conditions. As a result, the ratio $\lambda_p^{DLG}/\lambda_p^{SLG} \approx (\sqrt{n_{top}} + \sqrt{n_{bot}})/\sqrt{n_{top} + n_{bot}}$ falls within $[1, \sqrt{2}]$ [4].

In order to fit the experimental data, we calculate the optical conductivity of bilayer graphene $\sigma_{BLG}(\omega)$ using a tight binding model as described in detail in Ref. 5. In such a model, we considered several parameters in the model: $\gamma_0$, $\gamma$ and $\delta n_0$. Here, $\gamma_0$ is the in-plane hopping energy between nearest-neighbor atoms, $\gamma$ (labeled as $\gamma_1$ in Ref. 5 and others) is the interlayer hopping/tunneling energy between the sublattices A and B of the two graphene layers (Figure 1d in the main text), and $\delta n_0$ describes asymmetric doping between the top and bottom graphene layers. For simplicity, we didn't consider higher order hopping terms such as $\gamma_3$ and $\gamma_4$ that cause trigonal warping and electron-hole asymmetry, respectively. According to previous studies[5,6], the higher order terms only have notable effects at low energy and low temperature. Following Ref. 5, we set $\gamma_0$ and $\gamma$ to be 3 eV and 0.4 eV respectively in the band structure calculation. The parameter $\delta n_0$ is a sample-dependent parameter that can be fitted by our modeling[5]. Based on the calculated band structure, we then computed the optical conductivity of bilayer graphene using the Kubo formula that considers all the intraband and interband transitions in bilayer graphene. As shown in Figure 3a, our method is sufficient to reproduce quantitatively well the experimental data.

In Figure S5 below and Figure 4 of the main text, we show the voltage- and frequency-dependent color maps for calculated real ($\sigma_1$) and imaginary ($\sigma_2$) parts of optical conductivities for single-layer graphene, bilayer graphene (BLG-1 & BLG-2) and double-layer graphene (DLG-1 & DLG-2). As introduced in the main text, the asymmetric doping parameter $\delta n_0$ is fitted to be $2.5 \times 10^{11}$ cm$^{-2}$ and $9.1 \times 10^{11}$ cm$^{-2}$ for BLG-1 (DLG-1) and BLG-2 (DLG-2), respectively. In order to facilitate comparison, Figure S5 and Figure 4 (main text) are plotted under the same color scale. By comparing Figure S5 and Figure 4, one can see that in most parts of the plasmon on regions ($\sigma_2 > 0$), $\sigma_1$ is close to zero, so we have $\sigma_2 \gg \sigma_1$, which is the criterial for formation of well-defined plasmons. $\sigma_1$ becomes larger than $\sigma_2$ only at low frequency close to the DC limit or the plasmon off region ($\sigma_2 \leq 0$).

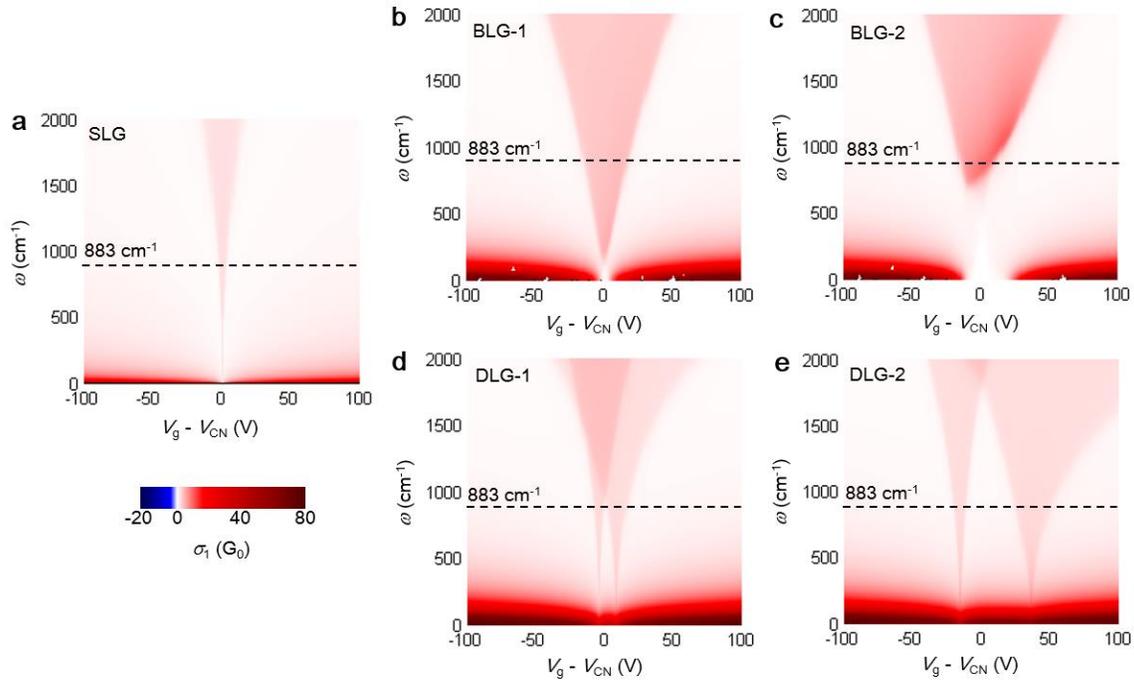

**Figure S5.** Color maps of the real-part of optical conductivity. **a-e**, Calculated voltage- and frequency- dependent maps of the real part of the optical conductivity $\sigma_1(V_g\text{-}V_{CN}, \omega)$ of SLG, BLG-1, DLG-1, BLG-2, DLG-2. Here, BLG-1 and BLG-2 are Bernal-type bilayer graphene samples with $\delta n_0 = 2.5 \times 10^{11}$ cm$^{-2}$ and $\delta n_0 = 9.1 \times 10^{11}$ cm$^{-2}$, respectively. DLG-1 and DLG-2 are randomly-stacked double-layer graphene samples with $\delta n_0 = 2.5 \times 10^{11}$ cm$^{-2}$ and $\delta n_0 = 9.1 \times 10^{11}$ cm$^{-2}$, respectively. The conductivity unit

is $G_0 = e^2/4\hbar$.

## 5. Plasmon dispersion

In Figure S6, we plot the frequency ($\omega$) – momentum ($q$) dispersion diagrams for all the surface modes in single-layer, bilayer and double-layer graphene on SiO$_2$/Si substrates. These dispersion plots are calculated by evaluating numerically the imaginary part of the reflection coefficient Im($r_p$) for the entire sample/substrate system using the transfer matrix method[3]. We set the gate voltage to be $V_g$-$V_{CN}$ = 15 V, where surface plasmons in both BLG-1 and BLG-2 are turned off at $\omega_{IR}$ = 883 cm$^{-1}$ (Figure 3b), but only BLG-2 is on the insulating state. As introduced in detail in the main text, at the insulating state, plasmons of bilayer graphene are turned off completely at all frequencies. The bright curves in these diagrams are dispersion curves for various surface modes. The relatively flat mode above $\omega$ = 1100 cm$^{-1}$ is the surface phonon mode of SiO$_2$, while the mode below following a $q^{1/2}$ scaling is the surface plasmon mode. When these two modes approach each other, an anti-crossing phenomenon occurs due to plasmon-phonon coupling[3]. In the current work, we mainly focused on the graphene plasmon mode. In order for us to launch and detect the plasmon mode, the plasmon dispersion curve should cross the horizontal dashed line, which is set at our excitation frequency $\omega_{IR}$ = 883 cm$^{-1}$ (Figure S6). Apparently, the SLG, DLG-1 and DLG-2 plasmon modes crosses our excitation frequency (horizontal dashed lines), and can therefore be excited. For BLG-1, the plasmon mode appears at lower frequencies, precluding plasmon excitation at $\omega_{IR}$ = 883 cm$^{-1}$. In the case of BLG-2 that is on the insulating state, no plasmon mode can be seen in the entire frequency range.

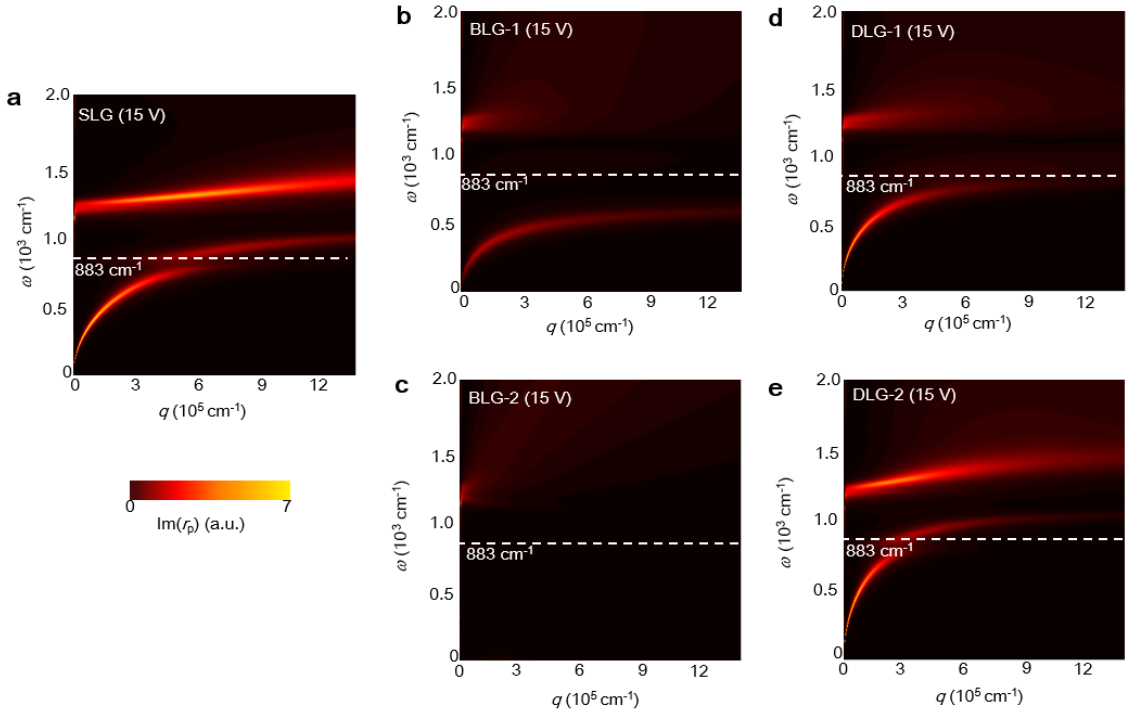

**Figure S6.** Dispersion diagrams. **a-e**, Calculated frequency ($\omega$) – momentum ($q$) dispersion diagrams for surface modes of SLG, BLG-1, DLG-1, BLG-2 and DLG-2 on SiO$_2$/Si substrate at $V_g$ - $V_{CN}$ = 15 V. Here we plot the imaginary part of the reflection coefficient Im($r_p$). BLG-1 and BLG-2 are Bernal-type bilayer graphene samples with $\delta n_0$ = 2.5 × 10$^{11}$ cm$^{-2}$ and $\delta n_0$ = 9.1 × 10$^{11}$ cm$^{-2}$, respectively. DLG-1 and DLG-2 are randomly-stacked double-layer graphene samples with $\delta n_0$ = 2.5 × 10$^{11}$ cm$^{-2}$ and $\delta n_0$ = 9.1 × 10$^{11}$ cm$^{-2}$, respectively**.** Horizontal dashed lines mark our excitation and probing frequency $\omega_{IR}$ = 883 cm$^{-1}$.

## 6. Calculation of the plasmon wavelength

In order to calculate the plasmon wavelength, we first determine the optical conductivity of the graphene layers using the methods as detailed above in Section 4, and then obtain the plasmon dispersion by evaluating numerically the imaginary part of the reflection coefficient Im($r_p$) for the entire sample/substrate system by using the transfer matrix method (Figure S6). Based on the dispersion diagram, we can determine directly the plasmon wavevector $q_p$ hence the plasmon wavelength $\lambda_p = 2\pi/q_p$. With this method, we produced the gate-dependent plasmon wavelength for all the graphene layers. In addition to numerical method, an analytical solution about the plasmon dispersion of a two-dimensional conducting film can be written as $q_p = i\omega\kappa(\omega)/2\pi\sigma(\omega)$ under the long-wavelength approximation[1]. Here $\kappa(\omega) = [1+\varepsilon_{sub}(\omega)]/2$ is the effective dielectric function for the environment of graphene at the interface between air and substrate, $\sigma(\omega) = \sigma_1(\omega) + i\sigma_2(\omega)$ is the optical conductivity of the film. According to the dispersion equation, in order to observe well-defined plasmons, the optical conductivity should be predominantly imaginary ($\sigma_2 \gg \sigma_1$). In this case, the plasmon wavelength $\lambda_p$ is roughly proportional to $\sigma_2$.

## 7. Further discussions about doping dependence of $\lambda_p$

In Figure 3a of the main text, we plot both the measured and calculated $\lambda_p$ versus gate voltages that is proportional to the carrier density. Here in Figure S7b, as a comparison, we change the horizontal axis from the scale of density to energy. The conversion is done based on the calculated dependence curves between the carrier density ($n$) and Fermi energy ($E_F$) (Figure S7a). From Figure S7a, one can see that $E_F \sim n^{1/2}$ for single-layer graphene with linear bands and $E_F \sim n$ for bilayer graphene with parabolic bands. The calculated $n$ - $E_F$ dependence curves of BLG-1 and BLG-2 with tight-binding model show a mixed behavior: $E_F \sim n$ (corresponding to parabolic dispersion) close to the charge neutrality point (CNP) and $E_F \sim n^{1/2}$ (corresponding to linear dispersion) at high doping regime. The step-like wiggle features close to the CNP of BLG-2 and also BLG-1 (not obvious in the current axis scale) are manifestations of the bandgap opening close to the CNP.

A number of new features can be seen when comparing Figure S7b with Figure 4 in the main text. First, we notice that when plotting $\lambda_p$ versus $E_F$ (Figure S7b), the plasmon-off regions of single-layer graphene (black curve) and gapless bilayer graphene (green curve) have the same size ($W$). This observation confirms that the plasmon-off regions of these two gapless systems share the same origin: interband transitions that starts at $\omega = 2E_F$ and causes Landau damping. Second, bandgap opening increases the size of the plasmon-off region, but the relative size change in energy units (Figure S7b) appears to be smaller compared to that in density units (Figure 4). This is related to the rapidly increasing density of states of bilayer graphene away from the CNP (Figure S7a): $\Delta n / \Delta E_F \sim n^i$, ($0 < i \leq 0.5$). Third, at high doping regime, $\lambda_p$ scales linearly with $E_F$ (Figure S7b) and the slopes of the $\lambda_p$ curves reflect the dispersion of the modeled bands. Apparently, bilayer graphene with parabolic bands (green curve) have a more rapidly growing $\lambda_p$ than that of single-layer graphene with linear bands (black curve). Realistic bilayer graphene, such as BLG-1 (blue curve and dots) and BLG-2 (red curve and dots), resides between the green and black curves. Finally, at high doping regime where $E_F$ is far above the bandgap, $\lambda_p$ at a given $E_F$ shows weak dependence with the size of the bandgap. As a result, the data points of BLG-1 and BLG-2 in Figure S7b overlap with each other away from the CNP. In contrast, bandgap effects can be seen more clearly in Figure 4 of the main text where we plot the $\lambda_p$ versus gate voltages. This is because more carriers have been injected to fill the low-energy bands with gap opening, and consequently, at a fixed carrier density (or gate voltage), $E_F$ of bilayer graphene with bigger bandgap (e.g. BLG-2) is lower than that with smaller bandgap (e.g. BLG-1) as shown clearly in Figure S7a.

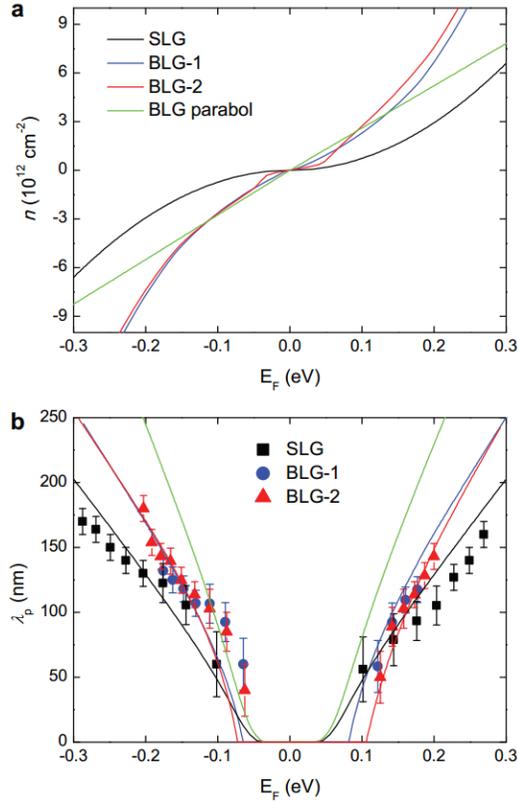

**Figure S7. a**, Calculated carrier density ($n$) versus Fermi energy ($E_F$) of single-layer graphene (black), parabolic-band bilayer graphene (green), and of BLG-1 & BLG-2 from the tight-binding model. **b**, Experimental and modeling plasmon wavelength ($\lambda_p$) versus Fermi energy ($E_F$). The black, blue and red data points are experimentally measured $\lambda_p$ for single-layer graphene (SLG) and two bilayer graphene samples (BLG-1 and BLG-2), respectively. The blue and red curves are calculated plasmon wavelength for BLG-1 and BLG-2 with the tight-binding model. The black and green curves are calculated plasmon wavelength for SLG and bilayer graphene with simplified parabolic-band model.